\documentclass[pra, showpacs, twocolumn, a4paper, nofootinbib]{revtex4}
\usepackage{graphicx}
\usepackage{amsmath, amsfonts, amssymb, bm}
\usepackage{amsmath}
\usepackage{verbatim}
\usepackage{bm}

\usepackage{graphicx,psfrag,bm,times,amsmath,amssymb}

\begin{document}

\title{Thresholdless dressed-atom laser in a photonic band-gap material}

\author{Gao-xiang \surname{Li}$^{a}$}
\email{gaox@phy.ccnu.edu.cn}
\author{Min \surname{Luo}$^{a}$}
\author{Zbigniew \surname{Ficek}$^{b}$}
\affiliation{$^{a}$Department of Physics, Huazhong Normal University, Wuhan 430079, China\\
$^{b}$Department of Physics, School of Physical Sciences, The University of Queensland,
Brisbane, Australia 4072}

\date{\today}

\begin{abstract}
We demonstrate the capability of complete thresholdless lasing
operation between dressed states of a two-level atom located inside
a microscopic cavity engineered in a photonic band-gap material. We
distinguish between threshold and thresholdless behaviors by
analyzing the Mandel's $Q$ parameter for the cavity field. We find
that the threshold behavior depends on whether the spontaneous
emission is or is not present on the lasing transition. In the
presence of the spontaneous emission, the mean photon number of the
cavity field exhibits threshold behavior indicating that the system
may operate as an ordinary laser. When the spontaneous emission is
eliminated on the lasing transition, no threshold is observed for
all values of the pumping rate indicating the system becomes a
thresholdless laser. Moreover, we find that under a thresholdless
operation, the mean photon number can increase nonlinearly with the
pumping rate, and this process is accompanied by a sub-Poisson
statistics of the field. This suggests that the nonclassical
statistics can be used to distinguish a nonlinear operation of the
dressed-atom laser.
\end{abstract}


\maketitle

\section{Introduction}\label{sec1}

\ \  A single atom laser, where no more than one atom is present in an optical resonator, is a fundamental model for study low intensity lasing operations and has been a subject of great interest over the
years~\cite{ms92,gb93,hg95,lg97,jg99,bb04,kk08}.
Traditional laser theories developed over fifty years consider an ensemble of multi-level atoms, a macroscopic medium incoherently pumped inside the resonator~\cite{sl67,sz97}.
It has been recognized that crucial for the laser operation is the presence of spontaneous emission, and therefore the laser theories are essentially divided into two categories, the so-called class-$A$ lasers that do not include spontaneous emission between the lasing levels, and class-$B$ lasers that take into account all the spontaneous emission in the system. The results have predicted differences between these two classes in the threshold behavior and output power. Particularly interesting is the prediction of a low threshold behavior of the class-$B$ lasers which makes them close to operate as thresholdless devices.

\ \ At present, research in the area of the laser theory is mostly focused on single-atom lasers and their practical applications in photonics, nano and quantum technologies~\cite{mck,kz07,lc07}.
Although a single-atom laser is admittedly an elementary model, it has the advantage over a multi-atom laser that in any practical realization of the laser one is not concerned with many difficulties such as fluctuations of the number of atoms. In addition, single-atom lasers operate at very low threshold, close to "thresholdless" lasing, require very low pumping rates and produce a low intensity light with a small number of photons~\cite{rc94,dw02}.
These characteristics require the quantum rather than a semiclassical description of the laser. Therefore, it is not a surprise that  many quantum features of single-atom lasers, such as sub-Poissonian photon statistics, photon antibunching, squeezing and vacuum Rabi splitting have been predicted and experimentally observed~\cite{mck,tp}.

\ \ In this paper, we present a very simple and practical model of a
dressed-atom laser incoherently pumped by spontaneous emission. We
show that the system can operate as a linear or nonlinear {\it
thresholdless} laser at all values of the pumping rate. The earliest
work demonstrated~\cite{rc94,dm88,bk94,pd99} that a thresholdless operation can be achieved
when the fraction of spontaneous emission into the cavity (lasing) mode, determined by
the so-called $\beta$ parameter approaches the limit $\beta\rightarrow 1$. In other words, in order to observe thresholdless lasing all of the spontaneous emission must be directed to the laser mode.
This is a significant practical problem requiring a very strong coupling of the atom to the cavity mode and a very small size of the cavity. The mechanism considered here for producing thresholdless lasing is quite different from that considered before. Essentially, we show that the system can work as a thresholdless laser when the spontaneous emission is {\it eliminated} rather than being focused on the lasing transition. To demonstrate this, we go beyond
the traditional approach by first driving the atom with a strong
laser field and next couple the resulting dress-atom system to a
frequency dependent reservoir, which is engineered by a photonic
band-gap (PBG) material~\cite{lf}. This sequence in which the
interactions are handled leads to a possibility of a  dynamical
switching of the system from the class-$B$ to class-$A$ laser and
vice versa. These effects result from the filtering property of the
photonic crystal that effectively forms a frequency dependent
reservoir of a specific spectral function (step function) of the
radiation modes. A photonic crystal that is a periodic dielectric
structure, can prohibit light propagation over a continuous range of
frequencies, irrespective of the direction of
propagation~\cite{p1,p2}. It is also known that in photonic crystals
extremely small micro-cavity mode volumes and very high cavity
factors  can be realized~\cite{p2,gr,op,sj,aa,bh}.  It is associated with
the unique properties of photonic crystals, i.e., the photonic
density of states within or near a photonic band gap can almost
vanish or exhibit discontinuous changes as a function of frequency
with appropriate engineering, which is essentially different from
its free-space counterpart. Ultralow lasing thresholds in the
presence of a photonic crystal have been predicted~\cite{sh99} and
observed~\cite{sh06,ea07} in several different high-quality
nanocavities.

\begin{figure}[hbp]
\includegraphics[width=3cm,keepaspectratio,clip]{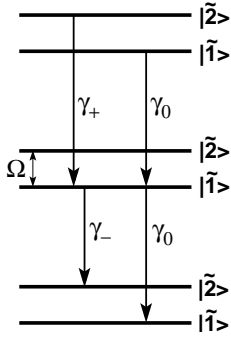}
\caption{Energy levels of the dressed atom system with the transition rates to and out of the dressed state $|\tilde{1}\rangle$. The transition rate $\gamma_{-}$ corresponds to spontaneous emission rate at the lasing frequency $\omega_{L}-2\Omega$.}
\label{fig1a}
\end{figure}

\section{The Model}\label{sec2}

\ \ We study threshold properties of the cavity field generated from
a dressed-atom system located inside a microscopic cavity of
frequency $\omega_{c}$ engineered inside a PBG material. The
considered dressed states are the eigenstates of a single two-level
atom composed of two energy states $|1\rangle$ and $|2\rangle$,
separated by transition frequency $\omega_{a}$, and driven by a
strong coherent laser field of a frequency $\omega_L$ and the
resonant Rabi frequency $\epsilon$. The dressed states are of the
form~\cite{sb87}
\begin{eqnarray}
|\tilde{1}\rangle &=& (\cos\phi)|1\rangle+(\sin\phi)|2\rangle,\nonumber\\
|\tilde{2}\rangle &=&  (\sin\phi)|1\rangle-(\cos\phi)|2\rangle ,\label{e1}
\end{eqnarray}
where $ \cos^{2}\phi = (1+\Delta_{a}/2\Omega)/2$,
$\Delta_{a}=\omega_{a}-\omega_{L}$ describes  detuning of the laser
frequency from the atomic resonance
and~$2\Omega=(4\epsilon^{2}+\Delta_{a}^{2})^{1/2}$ is the Rabi
frequency of the detuned field.

\ \ We couple the dressed states to a single mode of a high-$Q$ microcavity engineered within a photonic crystal and to the photonic crystal vacuum modes. The coupling is determined by the Hamiltonian which under the electric dipole and rotating-wave approximations has the form
\begin{eqnarray}
H_{bg} &=& i\hbar{g}(a^{\dag}\sigma_{12}-\sigma_{21}a) \nonumber \\
&& +i\hbar\sum_{\lambda}g_{\lambda}(\omega_{\lambda})
(a^{\dag}_{\lambda}\sigma_{12}-\sigma_{21}a_{\lambda})  ,\label{e2}
\end{eqnarray}
where $a\ (a^{\dag})$ and $a_{\lambda}\ (a^{\dag}_{\lambda})$ are the cavity mode and the photonic crystal vacuum reservoir annihilation (creation) operators, respectively.  The coefficient $g$ describes the strength of the coupling between the atom and the cavity mode that we assume to be constant
independent of frequency,  and $g_{\lambda}(\omega_{\lambda})$ describes the strength of the coupling between the atom and the vacuum  modes of the photonic crystal. In practice this model can be realized by embedding an atom in a dielectric microcavity (defect) placed within a two-mode waveguide channel in a 2D PBG microchip~\cite{lf}.

\ \ The coupling constant $g_{\lambda}(\omega_{\lambda})$ contains the information about the frequency dependent mode structure of the photonic crystal and, in general, can be written as
\begin{eqnarray}
g_{\lambda}(\omega_{\lambda}) = g_{\lambda}D(\omega_{\lambda}) ,\label{e3}
\end{eqnarray}
where $g_{\lambda}$ is a constant proportional to the dipole moment of the atom, and
$D(\omega_{\lambda})$ is the transfer function of the reservoir which, for a photonic crystal, is in the form of the unit step function, $|D(\omega_{\lambda})|^{2} = u(\omega_{\lambda}-\omega_{b})$,
where $\omega_{b}$ is the photonic density of states band edge frequency.
Thus, $|D(\omega_{\lambda})|^{2} =0$ for $\omega_{\lambda}<\omega_{b}$ and
$|D(\omega_{\lambda})|^{2} =1$ for $\omega_{\lambda}>\omega_{b}$. In a real band gap material, the band edge is not exactly in the form of the step function~\cite{lf}, but is rather in a form of a slop whose the width $\Delta\omega/\omega_{b}\approx 10^{-4}$, that is much smaller than the spontaneous emission rate $\gamma$ of the atomic bare transition and the cavity damping rate $\kappa$.
Since the transitions between the dressed states occur at three frequencies, $\omega_{L}$,
$\omega_{\pm}=\omega_{L}\pm 2\Omega$, and the frequencies $\omega_{\lambda}<\omega_{b}$ are forbidden in the band gap material, it is possible to eliminate spontaneous emission at selected frequencies of the dressed-atom system by a strong driving with $\Omega \gg \gamma,\kappa$.

\ \ The dynamics of the dressed-atom-cavity system coupled to the vacuum reservoir are determined by the master equation of the reduced density operator of the system.
In a frame oscillating at the frequency $\omega_{L}$, and under the secular approximation of
$\Omega \gg \gamma,\kappa$, where we ignore rapidly oscillating terms at frequencies $2\Omega$ and $4\Omega$, the master equation is of the form~\cite{lf,qf93}
\begin{eqnarray}
\partial{\rho}/\partial{t} & =&  \frac{1}{2}g\sin (2\phi) \ [a^{\dag}R_{3}e^{i\Delta_{c}t}
-aR_{3}e^{-i\Delta_{c}t},\rho]\nonumber\\
&&+g(\cos^{2}\phi) \ [a^{\dag}R_{12}e^{i(\Delta_{c}-2\Omega)t}-{\rm H.c.},\rho]\nonumber\\
&&-g(\sin^{2}\phi) \ [a^{\dag}R_{21}e^{i(\Delta_{c}+2\Omega)t} -{\rm H.c.},\rho]  \nonumber \\
&&+\frac{1}{8}\gamma_{0}(2R_3\rho{R_3}-\rho{R_3^2}-R_3^2\rho) \nonumber\\
&&+\frac{1}{2}\gamma_{-}(2R_{21}\rho{R_{12}}-R_{12}R_{21}\rho-\rho{R_{12}R_{21}}) \nonumber \\
&&+\frac{1}{2}\gamma_{+}(2R_{12}\rho{R_{21}}-R_{21}R_{12}\rho-\rho{R_{21}R_{12}}) \nonumber \\
&&+ \frac{1}{2}\kappa \left(2a\rho{a}^\dag-{a}^\dag{a}\rho-\rho{a}^\dag{a}\right) ,\label{e4}
\end{eqnarray}
where $R_{ij}=|\tilde{i}\rangle\langle \tilde{j}|$ are the dressed-state transition operators,
$R_{3}=|\tilde{2}\rangle\langle \tilde{2}|-|\tilde{1}\rangle\langle \tilde{1}|$ is the population difference operator, $\Delta_{c}=\omega_{c}-\omega_{L}$ is the detuning of the laser frequency from the cavity frequency, and
\begin{eqnarray}
\gamma_{0} &=& \gamma \sin^{2}(2\phi) \ u(\omega_{L}-\omega_{b}) ,\ \
\gamma_{-}\!=\!\gamma (\sin^{4}\!\phi) u(\omega_{-}-\omega_{b})  ,\nonumber \\
\gamma_{+} &=& \gamma (\cos^{4}\phi) u(\omega_{+} -\omega_{b}) ,\label{e5}
\end{eqnarray}
are the damping rates between the dressed states of the system. The coefficient $\gamma_{0}$ corresponds to spontaneous emission occurring at two transitions
of the dressed atom; One from the lower dressed state $|\tilde{1}\rangle $ to the lower  dressed state
of the manifold below and the other from the upper dressed state $|\tilde{2}\rangle $ to the upper dressed state of the manifold below. These transitions occur at frequency $\omega_{L}$.
The coefficient $\gamma_{+}$ corresponds to spontaneous emission from the upper dressed
state to the lower  dressed state of the manifold below and occurs at frequency
$\omega_{+}=\omega_{L}+2\Omega$, whereas the coefficient $\gamma_{-}$ corresponds to spontaneous emission from the lower dressed state to the upper dressed state of the manifold below and occurs at frequency $\omega_{-}=\omega_{L}-2\Omega$.

\ \ We note here that the damping rate $\gamma_{-}$ appears as the decay rate between the laser levels, whereas the rate~$\gamma_{+}$, that is usually recognized as a damping rate plays, in fact, the role of an incoherent pumping of the dressed system from $|\tilde{2}\rangle$ to $|\tilde{1}\rangle$. In other words, it is a pure incoherent pumping process that transfers the population to the upper state of the lasing transition. We should point out, and as it is pictured in Fig.~\ref{fig1a}, the state $|\tilde{1}\rangle$ is also populated with the rate~$\gamma_{0}$. However, the state is also depopulated with the same rate, so the spontaneous emission with the rate $\gamma_{+}$ is the only pumping mechanism in the dressed-atom system. It is important to emphasize here that despite the presence of the external laser field the lasing transition is pumped by a pure incoherent process. The coherent field appears here as a "dressing" field of the atom. Therefore, the process investigated in this paper is {\it not} the resonant scattering of coherent field by the atom placed in a photonic band gap material.

\ \ The master equation (\ref{e4}) is general in terms of the detunings so that both the atom and the cavity field can be either on resonance or off resonance with the laser field. In the following, we specify the frequency of the cavity field to be tuned to exact resonance with the low-frequency Rabi sideband,
i.e. $\Delta_{c}=-2\Omega$ that will serve as the lasing frequency.
Under this specific tuning, the spontaneous emission at this frequency is cancelled in the full band-gap situation, $\gamma_{-}=0$, but the spontaneous emission on the other transition can be allowed or cancelled depending on the frequencies of the transition relative to the bang edge frequency. In the absence of the band gap material, spontaneous emission is allowed at all the dressed atom frequencies.

\section{Dressed-Atom Laser}\label{sec3}

\ \ We study the lasing properties of the cavity field by evaluating the mean number of photons
$\langle n\rangle$ in the cavity field and the Mandel's $Q$ parameter determining the statistics of the field. The Mandel's $Q$ parameter is related to the Fano factor $F=\langle(\Delta n)^{2}\rangle/\langle n\rangle$, commonly used to identify threshold behavior of the cavity field, through the
relation~$Q=F-1$. In addition, the parameter $Q$ is used here to quantify the cavity field as a "laser" field. Simply, when $\langle n\rangle\neq 0$ and $Q=1$ the cavity field is in a coherent state.

\ \ We expand the density operator of the system in terms of the atomic states and the photon number states of the cavity field, and find that $\langle n\rangle$ and $\langle n^{2}\rangle$ defined as
\begin{eqnarray}
\langle n\rangle =\sum_{n=0}^{\infty} nP_{n}^{(1)} ,\quad
\langle n^{2}\rangle =\sum_{n=0}^{\infty} n^{2}P_{n}^{(1)} ,\label{e6}
\end{eqnarray}
are determined by the photon-number distribution function $P_{n}^{(1)}=\langle n| \rho_{1}|n\rangle$,
where $\rho_{1} = {\rm Tr}_{a}(\rho) = \rho_{\tilde{1}\tilde{1}} +\rho_{\tilde{2}\tilde{2}}$
is the reduced density operator of the cavity field, and $\rho_{\tilde{1}\tilde{1}}$
and $\rho_{\tilde{2}\tilde{2}}$ are the populations of the dressed states of the system.

\ \ According to Eq.~(\ref{e6}), what we need is only to find the distribution function $P_{n}^{(1)}$ to completely determine the statistics of the cavity field from which one can distinguish threshold and thresholdless behaviors of the dressed-atom laser. In order to find the distribution function, we introduce Hermitian combinations of the density matrix elements
\begin{align}
\rho_{1}&=\rho_{22}+\rho_{11} ,\quad \rho_{2} =\rho_{22}-\rho_{11}  ,\nonumber\\
\rho_{3}&=(a^\dag\rho_{12} +\rho_{21}a)/2 ,\quad
\rho_{4}=(a\rho_{21}+\rho_{12}a^\dag)/2 ,\label{e7}
\end{align}
and find that in the photon number representation they satisfy the following equations of motion
\begin{align}
\dot{P}_{n}^{(1)}=& -\kappa{n}P_{n}^{(1)}+\kappa(n+1)P_{n+1}^{(1)} -2g_{1}(P_{n}^{(3)}-P_{n}^{(4)}) , \nonumber \\
\dot{P}_{n}^{(2)}=&-(\gamma_{+}+\gamma_{-}+\kappa n)P_{n}^{(2)}
-(\gamma_{+}-\gamma_{-})P_{n}^{(1)}\nonumber\\
&+\kappa(n+1)P_{n+1}^{(2)} -2g_{1}(P_{n}^{(3)}+P_{n}^{(4)}) ,\nonumber \\
\dot{P}_{n}^{(3)}=& -\frac{1}{2}\left[4\gamma_{0} +\gamma_{+} +\gamma_{-}
+\kappa\left(2n-1\right)\right]P_{n}^{(3)} \nonumber \\
&+\frac{1}{2}n g_{1}\left(P_{n}^{(1)}-P_{n-1}^{(1)}+P_{n-1}^{(2)}+P_{n}^{(2)}\right) \nonumber \\
&+\kappa(n+1)P_{n+1}^{(3)}-\kappa P_n^{(4)}  ,\nonumber \\
\dot{P}_{n}^{(4)}=& -\frac{1}{2}\left[4\gamma_{0} +\gamma_{+} +\gamma_{-}
+\kappa\left(2n+1\right)\right]P_{n}^{(4)} \nonumber \\
&+\frac{1}{2}(n+1)g_{1}\left(P_{n+1}^{(1)}-P_{n}^{(1)}+P_{n+1}^{(2)}+P_{n}^{(2)}\right) \nonumber  \\
&+\kappa(n+1)P_{n+1}^{(4)}  ,\label{e8}
\end{align}
where $g_{1}=g\sin^{2}\!\phi$ is the "effective" coupling constant of the cavity field to the dressed-atom
transition $|\tilde{1}\rangle \rightarrow |\tilde{2}\rangle$.

\begin{figure}[hbp]
\includegraphics[width=\columnwidth,keepaspectratio,clip]{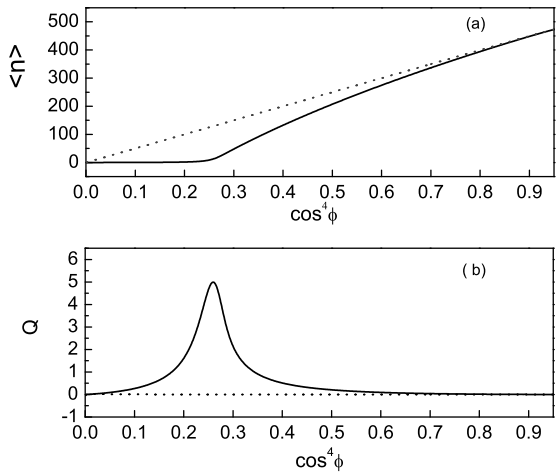}
\caption{(a) Mean photon number $\langle n\rangle$ and (b) Mandel's
$Q$ parameter as a function of the pumping rate $\cos^{4}\phi
=\gamma_{+}/\gamma$ for $\kappa/g =10^{-4}$ and $\kappa/\gamma
=10^{-3}$ with $(\gamma_{-}=\gamma_+$, solid line) and without
$(\gamma_{-}=0$, doted line) spontaneous emission on the lasing
transition.} \label{fig1}
\end{figure}

\ \ In the following, we consider the stationary and spectral properties of the cavity field.  and therefore, we solve the set of equations~(\ref{e8}) numerically for the steady-state value of the distribution function $P_{n}^{(1)}$ and calculate the mean number of photons and the Mandel's $Q$ parameter.

\ \ Figure~\ref{fig1} illustrates the mean photon number and the Mandel's $Q$ parameter versus the pumping rate $\gamma_{+}$. The plots in Fig.~\ref{fig1}(a) show a threshold behave in the presence of the spontaneous emission on the lasing transition, and no threshold when the spontaneous emission is completely suppressed.
The threshold actually occurs at the pumping rate attaining  the value $\cos^{4}\phi =0.25$ corresponding to the population inversion between the dressed states of the lasing transition. More precisely, we do not see a sharp threshold behave but rather a kink in the mean photon number for no band-gap situation which we can account as an indication of the presence of laser threshold. However, we can find that the kink becomes less visible as the ratio $\kappa/\gamma_{+}$ increases, approaching the value corresponding to the thresholdless lasing occurring at a full band-gap situation. Therefore, one could argue that a kink is not a clear indication of threshold behavior.

\ \ To resolve this problem, we study the statistical properties of
the cavity field by using the Mandel's $Q$ parameter as the
signature of threshold behavior. The disappearance of the threshold
is manifested by the change of the photon statistics from
super-Poissonian to Poissonian. This is shown in Fig.~\ref{fig1}(b),
where we plot the Mandel's $Q$ parameter as a function of the
pumping rate. Evidently, a sharp change of the photon statistics
from super-Poissonian to Poissonian at $\cos^{4}\phi =0.25$ is a clear
indication of a threshold behave for no band-gap situation~\cite{pr94,bm96}

\begin{figure}[hbp]
\includegraphics[width=\columnwidth,keepaspectratio,clip]{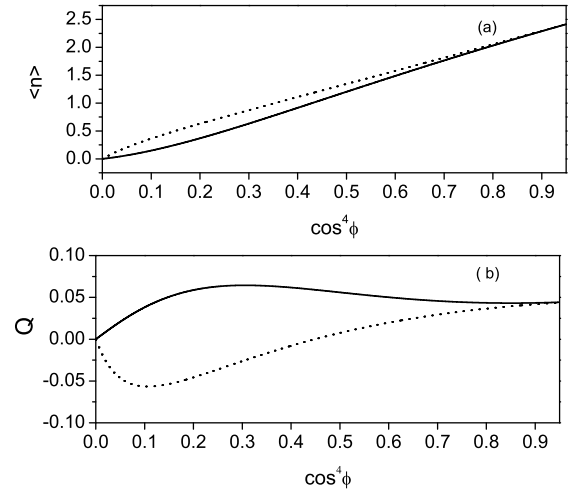}
\caption{(a) Mean photon number $\langle n\rangle$ and (b) Mandel's
$Q$ parameter as a function of the pumping rate $\cos^{4}\phi
=\gamma_{+}/\gamma$ for $\kappa/g =10^{-4}$ and  $\kappa/\gamma
=0.2$ with $(\gamma_{-}=\gamma_+$, solid line) and without
$(\gamma_{-}=0$, doted line) spontaneous emission on the lasing
transition.} \label{fig2}
\end{figure}

\ \ Poisson statistics of the field is traditionally recognized as a
signature of a lasing operation. We find an interesting effect that
under complete thresholdless operation, the mean photon number can
increase nonlinearly with the pumping rate and this process is
accompanied by a sub-Poisson statistics of the cavity field. Thus,
under thresholdless situation, the dressed laser can operate as a
nonlinear laser and a sub-Poisson statistics of the cavity field can
be recognized as the signature of the nonlinear operation. This is
shown in Fig.~\ref{fig2}, where we plot the mean photon number and
the $Q$ parameter for $\kappa/g\ll 1$ but not too small
$\kappa/\gamma$, i.e. $\kappa/\gamma<1$. It is evident from the
figure that in the limit of a low pumping rate, the mean photon
number evolves nonlinearly with the pumping rate and this nonlinear
evolution is accompanied by a sub-Poissonian statistics of the
cavity field.

\ \ In order to get a better insight into the threshold and thresholdless properties of the dressed-atom system, we take the limit of the pumping rate much larger than the cavity
losses, $\gamma_{+}\gg \kappa$. In this case, we find an explicit analytical
expression for the photon-number  distribution function of the
cavity field. We obtain the distribution function by setting the
derivatives to zero in Eq.~(\ref{e8}) and, after straightforward calculations, we~find
\begin{eqnarray}
P_{n}^{(1)} = \frac{1}{{\cal F}(1,m+1;\alpha)}\ \frac{\alpha^{n}m!}{(n+m)!} ,\label{e9}
\end{eqnarray}
where $\alpha = \gamma_{+}/2\kappa$,
\begin{eqnarray}
m = \frac{1}{2}\left[1\!+\!\frac{\gamma_{-}}{\kappa}\!+\!\frac{1}{4g_{1}^{2}}\left(4\gamma_{0}+\gamma_{+}
+\gamma_{-}\right)(\gamma_{+}\!+\!\gamma_-)\right] ,\label{e10}
\end{eqnarray}
and ${\cal F}(1,m+1;\alpha)$ is the confluent hypergeometric function.

\ \ In the limits of $\kappa/g\ll 1$ and $\kappa/\gamma_{+} \ll 1$, that correspond to the situation
presented in Fig.~\ref{fig1},  we find approximate expressions for the mean photon number
\begin{equation}
\langle n \rangle\approx\left(\gamma_+-\gamma_{-}\right)/(2\kappa) ,
\end{equation}
and for the Mandel's $Q$ parameter
\begin{equation}
Q=(\gamma_{-}+\kappa)/\gamma_{+} .
\end{equation}
These simple expressions show explicitly differences in threshold behave of the dressed-atom laser with and without a photonic band-gap material. Without the photonic band gap, $\gamma_{-}\neq 0$, and evidently there is a threshold at $\gamma_{+} = \gamma_{-}$, i.e.,
at $\cos^4\phi =1/4$. With the band-gap material, $\gamma_{-}=0$, and consequently no threshold  behavior is present.  Accordingly, the photon statistics is super-Poissonian, $Q>0$, when $\gamma_{-}\neq 0$, and becomes Poissonian, $Q\approx 0$, when $\gamma_{-}=0$.
Thus, the above approximate analytical expressions confirm the numerical analysis presented in
Fig.~\ref{fig1} that in the full band-gap situation of $\gamma_{-}=0$, the mean photon number behaves linearly with the pumping rate and the statistics is Poissionian.

\ \ We next use the distribution function  to examine the mean
photon number and photon statistics when the ratio $\kappa/\gamma $
is not too small, i.e. for $\kappa/\gamma <1$, the case illustrated
in Fig.~\ref{fig2}. For $\kappa/\gamma <1$, and in the limit of low
pumping rates, $\cos^4\phi <1/4$,  straightforward calculations
reveal that with the full band-gap situation of $\gamma_{-}=0$:
\begin{equation}
\langle n \rangle = 2\alpha(1-2\alpha) ,
\end{equation}
and accordingly the Mandel's $Q$ parameter takes the value
\begin{equation}
Q=-2\alpha/3 .
\end{equation}
Clearly, in this limit, the mean photon number has a quadratic dependence on the pumping rate and the photon statistics is sub-Poissonian, which confirms the prediction of the exact numerical results presented in Fig.~\ref{fig2}.

\ \ To confirm that our system behaviors like a typical laser, we calculate the stationary spectrum of the cavity field at the lasing frequency $\omega = \omega_{L}-2\Omega$. It is well known that there is no one-to-one correspondence between the spectral properties of the cavity field and its photon statistics~\cite{bm96}, but direct signature of lasing operation above the threshold is the decrease of the linewidth of the output cavity field below any linewidth in the system; the cavity linewidth and the linewidth of the lasing transition.
The stationary spectrum is defined as the Fourier transform of the two-time correlation function
$g(\tau) =\lim_{t\rightarrow \infty}\langle{a^{\dag}(t+\tau)a(t)}\rangle$ of the cavity field operators:
\begin{align}
S(\omega)=2 {\rm Re}\int^{\infty}_{0}{\rm e}^{i(\omega-\omega_{L}+2\Omega)\tau}g(\tau) d\tau .
\end{align}
 We calculate the spectrum numerically by Fourier transformation of the correlation
function $g(\tau)$ that is found by solving the master equation (\ref{e4}) in Fock representation for the steady state and subsequent evolution in time.

 \begin{figure}[hbp]
\includegraphics[width=\columnwidth,keepaspectratio,clip]{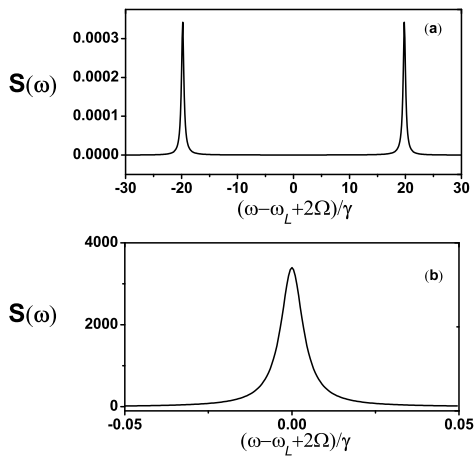}
\caption{The stationary spectrum of the output cavity field for $\kappa =0.05\gamma, g=20\gamma$ and two different values of the detuning of the driving laser field: (a)
 $\Delta_{a} =-10\epsilon $ and (b) $\Delta_{a} = 10\epsilon$.}
\label{fig3}
\end{figure}

Figure~\ref{fig3} shows the spectrum for $\kappa=0.05\gamma$,
$g=20\gamma$ and two, negative and positive, values of the laser detuning
$\Delta_{a}$ corresponding to a below and an above threshold value
of the pumping rate, respectively. Below the threshold, the spectrum
consists of two peaks located at frequencies $\pm g_1$ corresponding
to the vacuum Rabi splitting~\cite{smn,cb,zg,tr}.  Above the
threshold, the spectrum is composed of a single very narrow peak
located at the lasing frequency. The full width at half maximum
(FWHM) is about $0.005\gamma$ that is $1/10$ of the cavity decay
rate $\kappa$ and $1/200$ of the atomic decay rate $\gamma$. Thus,
we may conclude that the dressed-atom system behaves as a laser with
the output spectrum narrower than the cavity and the atomic linewidths.

\section{Conclusions}

 \ \ We have examined threshold behavior of a
dressed-atom laser engineered inside a photonic band-gap material
that appears as a frequency dependent reservoir for the dressed
system. We have found that by a suitable matching of the
dressed-atom transition frequencies to the guiding frequency of the
band gap material, it is possible to switch the dressed-atom laser
from a threshold to complete thresholdless operation. The results
show threshold behave in the presence of spontaneous emission on the
laser transition and thresholdless behave when the spontaneous
emission is suppressed. We have also found that under complete
thresholdless situation, the dressed laser can operate as a
nonlinear laser and this behave is indicated by a sub-Poisson
statistics of the cavity field. Further calculations of the spectrum of the cavity
field confirm that the dressed-atom system behaviors as a typical laser with
the spectral linewidth narrower than the cavity and atomic linewidths.

\section*{Acknowledgements}

We acknowledge financial support
from the National Natural Science Foundation of China (Grant Nos.
10674052 and 60878004), the Ministry of Education under project NCET
(grant no NCET-06-0671) and and SRFDP (under grant no.
200805110002), the National Basic Research Project of China (grant
no 2005 CB724508), and the Australian Research Council.


\begin{thebibliography}{99}
\bibliographystyle{unsrt}
\bibitem{ms92} Y. Mu and C. M. Savage,  Phys. Rev. A {\bf 46}, 5944 (1992).
\bibitem{gb93} C. Ginzel, H.-J. Briegel, U. Martini, B.-G. Englert, and A. Schenzle, Phys. Rev. A  {\bf 48}, 732 (1993).
\bibitem{hg95} P. Horak, K. M. Gheri, and H. Ritsch, Phys. Rev. A  {\bf 51}, 3257 (1995).
\bibitem{lg97} M. L\"offler, G. M. Meyer, and H. Walther, Phys. Rev. A {\bf 55}, 3923 (1997).
\bibitem{jg99} B. Jones, S. Ghose, J. P. Clemens, P. R. Rice, and L. M. Pedrotti, Phys. Rev. A  {\bf 60}, 3267 (1999).
\bibitem{bb04} A. D. Boozer, A. Boca, J. R. Buck, J. McKeever, and H. J. Kimble, Phys. Rev. A  {\bf 70}, 023814 (2004).
\bibitem{kk08} H.-J. Kim, A. H. Khosa, H. W. Lee, and M. S. Zubairy, Phys. Rev. A {\bf 77}, 023817 (2008).
\bibitem{sl67} M. Scully and W. E. Lamb, Jr., Phys. Rev. {\bf 159}, 208 (1967).
\bibitem{sz97} M. O. Scully and M. S. Zubairy, {\it Quantum Optics} (Cambridge University
Press, Cambridge, England, 1997).
\bibitem{mck} J. McKeever, A. Boca, A. D. Boozer, J. R. Buck, and H. J. Kimble,  Nature (London) {\bf 425}, 268 (2003).
\bibitem{kz07} M. Kiffner, M. S. Zubairy, J. Evers, and C. H. Keitel, Phys. Rev. A  {\bf 75}, 033816 (2007).
\bibitem{lc07} P. Lougovski, F. Casagrande, A. Lulli, and E. Solano, Phys. Rev. A {\bf 76}, 033802 (2007).
\bibitem{dw02} S. M. Dutra, J. P. Woerdman, J. Visser, and G. Nienhuis, Phys. Rev. A  {\bf 65}, 033824 (2002).
\bibitem{rc94} P. R. Rice and H. J. Carmichael, Phys. Rev. A  {\bf 50}, 4318 (1994).
\bibitem{tp} T. Pellizzari and H. Ritsch, Phys. Rev. Lett. {\bf 72}, 3973 (1994).
\bibitem{dm88} F. De Martini and G. R. Jacobovitz, Phys. Rev. Lett. {\bf 60}, 1711 (1988).
\bibitem{bk94} G. Bj\"ork, A. Karlsson, and Y. Yamamoto, Phys. Rev. A  {\bf 50}, 1675 (1994).
\bibitem{pd99} I. Protsenko, P. Domokos, V. Lefevre-Seguin, J. Hare, J. M. Raimond, and L. Davidovich, Phys. Rev. A  {\bf 59}, 1667 (1999).
\bibitem{lf} R. Wang and S. John, Phys. Rev. A {\bf 70}, 043805 (2004);
L. Florescu, S. John, T. Quang, and R. Wang, Phys. Rev. A {\bf 69}, 013816 (2004).
\bibitem{p1} E. Yablonovitch, Phys. Rev. Lett. {\bf 58}, 2059 (1987).
\bibitem{p2} S. John, Phys. Rev. Lett. {\bf 58}, 2486 (1987);
S. John and T. Quang, Phys. Rev. A {\bf 50}, 1764 (1994).
\bibitem{gr} A. D. Greentree, J. Salzman, S. Prawer, and L. C. L. Hollenberg,  Phys. Rev. A {\bf 73}, 013818 (2006).
\bibitem{op} O. Painter,  R. K. Lee, A. Scherer, A. Yariv, J. D. O'Brien, P. D. Dapkus, and I. Kim,
Science {\bf 284}, 1819 (1999).
\bibitem{sj} S. John, Phys. Rev. Lett. {\bf 53}, 2169 (1984).
\bibitem{aa} Y. Akahane, T. Asano, B.-S. Song, and S. Noda, Nature (London) {\bf 425}, 944 (2003).
\bibitem{bh} A. Badolato, K. Hennessy, M. Atature, J. Dreiser, E. Hu, P. M. Petroff and A. Imamoglu, Science {\bf 308}, 1158 (2005).
\bibitem{sh99} M. H. Szymanska, A. F. Hughs, and E. R. Pike, Phys. Rev. Lett. {\bf 83}, 69 (1999).
\bibitem{sh06} S. Strauf, K. Hennessy, M. T. Rakher, Y.-S. Choi, A. Badolato, L. C. Andreani, E. L. Hu, P. M. Petroff, and D. Bouwmeester, Phys. Rev. Lett. {\bf 96}, 127404 (2006).
\bibitem{ea07} D. Englund, H. Altug, J. Vuckovic, Appl. Phys. Lett. {\bf 91}, 071124 (2007).
\bibitem{sb87} C. Cohen-Tannoudji, J. Dupont-Roc, and G. Grynberg, {\it Atom-Photon
Interactions} (Wiley, New York, 1992), Chap. VI.
\bibitem{qf93} T. Quang and H. Freedhoff, Phys. Rev. A {\bf 47}, 2285 (1993).
\bibitem{pr94} T. Pellizzari and H. Ritsch, J. Mod. Opt. {\bf 41}, 609 (1994).
\bibitem{bm96} H.-J. Briegel, G. M. Meyer, and  B.-G. Englert, Phys. Rev. A  {\bf 53}, 1143 (1996).
\bibitem{smn} J. J. Sanchez-Mondragon, N. B. Narozhny, and J. H. Eberly, Phys. Rev. Lett. {\bf 51}, 550 (1983).
\bibitem{cb} H. J. Carmichael, R. J. Brecha, M. G. Raizen, H. J. Kimble, and P. R. Rice,
Phys. Rev. A {\bf 40}, 5516 (1989).
\bibitem{zg} Y. Zhu, D. J. Gauthier, S. E. Morin, Q. Wu, H. J. Carmichael, and T. W. Mossberg, Phys. Rev. Lett. {\bf 64}, 2499 (1990).
\bibitem{tr} R. J. Thompson, G. Rempe, and H. J. Kimble, Phys. Rev. Lett. {\bf 68}, 1132 (1992).
\end{thebibliography}
\end{document}